\documentclass[aps,prc,twocolumn,floatfix,showpacs,preprintnumbers,amsmath,amssymb,superscriptaddress]{revtex4-1}
\usepackage{hyperref}
\usepackage{latexsym}
\usepackage{verbatim}
\usepackage{graphics}
\usepackage{caption}
\usepackage{subfigure}
\usepackage{amsmath}
\usepackage{setspace}
\usepackage{graphicx,amsmath,color}
\usepackage{todonotes}
\usepackage{placeins}
\usepackage{afterpage}

\definecolor{lightred}{rgb}{1,0.5,0.5}
\definecolor{lightgreen}{rgb}{0.5,1,0.5}
\definecolor{darkgreen}{rgb}{0.5,0.9,0.5}

\def\pizT{$\pi^{0} \ $}

\definecolor{Mycolor2}{HTML}{0c9008}

\begin{document}

\title{Exclusive photoproduction of $\pi^0$ up to large values of Mandelstam variables $s, t$ and $u$ with CLAS}

\newcommand*{\ANL}{Argonne National Laboratory, Argonne, Illinois 60439, USA}
\newcommand*{\ANLindex}{1}
\affiliation{\ANL}
\newcommand*{\ASU}{Arizona State University, Tempe, AZ 85287, USA}
\newcommand*{\ASUIndex}{2}
\affiliation{\ASU}
\newcommand*{\BOCHUM}{Institut f\"ur Experimentalphysik I, Ruhr-Universit\"at 
	Bochum, 44780 Bochum, Germany}
	\newcommand*{\BOCHUMIndex}{3}
	\affiliation{\BOCHUM}
\newcommand*{\CSUDH}{California State University, Dominguez Hills, Carson, California 90747, USA}
\newcommand*{\CSUDHindex}{4}
\affiliation{\CSUDH}
\newcommand*{\CMU}{Carnegie Mellon University, Pittsburgh, Pennsylvania 15213, USA}
\newcommand*{\CMUindex}{5}
\affiliation{\CMU}
\newcommand*{\CUA}{The Catholic University of America, Washington, District of Columbia, 20064, USA}
\newcommand*{\CUAindex}{6}
\affiliation{\CUA}
\newcommand*{\SACLAY}{IRFU, CEA, Universit'e Paris-Saclay, F-91191 Gif-sur-Yvette, France}
\newcommand*{\SACLAYindex}{6}
\affiliation{\SACLAY}
\newcommand*{\CNU}{Christopher Newport University, Newport News, Virginia 23606, USA}
\newcommand*{\CNUindex}{7}
\affiliation{\CNU}
\newcommand*{\UCONN}{University of Connecticut, Storrs, Connecticut 06269, USA}
\newcommand*{\UCONNindex}{8}
\affiliation{\UCONN}
\newcommand*{\INDORE}{Indian Institute of Technology,  Simrol, Indore~453552, Madhya Pradesh, India}
\affiliation{\INDORE}
\newcommand*{\FU}{Fairfield University, Fairfield, Connecticut 06824, USA}
\newcommand*{\FUindex}{9}
\affiliation{\FU}
\newcommand*{\FERRARAU}{Universita' di Ferrara , 44121 Ferrara, Italy}
\newcommand*{\FERRARAUindex}{10}
\affiliation{\FERRARAU}
\newcommand*{\FIU}{Florida International University, Miami, Florida 33199, USA}
\newcommand*{\FIUindex}{11}
\affiliation{\FIU}
\newcommand*{\FSU}{Florida State University, Tallahassee, Florida 32306, USA}
\newcommand*{\FSUindex}{12}
\affiliation{\FSU}
\newcommand*{\Genova}{Universit$\grave{a}$ di Genova, 16146 Genova, Italy}
\newcommand*{\Genovaindex}{13}
\affiliation{\Genova}
\newcommand*{\GWUI}{The George Washington University, Washington, District of Columbia 20052, USA}
\newcommand*{\GWUIindex}{14}
\affiliation{\GWUI}
\newcommand*{\ISU}{Idaho State University, Pocatello, Idaho 83209, USA}
\newcommand*{\ISUindex}{15}
\affiliation{\ISU}
\newcommand*{\IKP}{Institut f\"ur Kernphysik, Forschungszentrum J\"ulich, 
	52424 J\"ulich, Germany}
	\newcommand*{\IKPIndex}{16}
	\affiliation{\IKP}
\newcommand*{\INFNFE}{INFN, Sezione di Ferrara, 44100 Ferrara, Italy}
\newcommand*{\INFNFEindex}{17}
\affiliation{\INFNFE}
\newcommand*{\INFNFR}{INFN, Laboratori Nazionali di Frascati, 00044 Frascati, Italy}
\newcommand*{\INFNFRindex}{18}
\affiliation{\INFNFR}
\newcommand*{\INFNGE}{INFN, Sezione di Genova, 16146 Genova, Italy}
\newcommand*{\INFNGEindex}{19}
\affiliation{\INFNGE}
\newcommand*{\INFNRO}{INFN, Sezione di Roma Tor Vergata, 00133 Rome, Italy}
\newcommand*{\INFNROindex}{20}
\affiliation{\INFNRO}
\newcommand*{\INFNTUR}{INFN, Sezione di Torino, 10125 Torino, Italy}
\newcommand*{\INFNTURindex}{21}
\affiliation{\INFNTUR}
\newcommand*{\ORSAY}{Institut de Physique Nucl\'eaire, CNRS/IN2P3 and Universit\'e Paris Sud, Orsay, France}
\newcommand*{\ORSAYindex}{22}
\affiliation{\ORSAY}
\newcommand*{\ITEP}{Institute of Theoretical and Experimental Physics, Moscow, 117259, Russia}
\newcommand*{\ITEPindex}{23}
\affiliation{\ITEP}
\newcommand*{\JMU}{James Madison University, Harrisonburg, Virginia 22807, USA}
\newcommand*{\JMUindex}{24}
\affiliation{\JMU}
\newcommand*{\KNU}{Kyungpook National University, Daegu 41566, Republic of Korea}
\newcommand*{\KNUindex}{25}
\affiliation{\KNU}
\newcommand*{\MISS}{Mississippi State University, Mississippi State, MS 39762-5167, USA}
\newcommand*{\MISSindex}{26}
\affiliation{\MISS}
\newcommand*{\UNH}{University of New Hampshire, Durham, New Hampshire 03824-3568, USA}
\newcommand*{\UNHindex}{27}
\affiliation{\UNH}
\newcommand*{\NSU}{Norfolk State University, Norfolk, Virginia 23504, USA}
\newcommand*{\NSUindex}{28}
\affiliation{\NSU}
\newcommand*{\OHIOU}{Ohio University, Athens, Ohio  45701, USA}
\newcommand*{\OHIOUindex}{29}
\affiliation{\OHIOU}
\newcommand*{\ODU}{Old Dominion University, Norfolk, Virginia 23529, USA}
\newcommand*{\ODUindex}{30}
\affiliation{\ODU}
\newcommand*{\RPI}{Rensselaer Polytechnic Institute, Troy, New York 12180-3590, USA}
\newcommand*{\RPIindex}{31}
\affiliation{\RPI}
\newcommand*{\URICH}{University of Richmond, Richmond, Virginia 23173, USA}
\newcommand*{\URICHindex}{32}
\affiliation{\URICH}
\newcommand*{\ROMAII}{Universita' di Roma Tor Vergata, 00133 Rome, Italy}
\newcommand*{\ROMAIIindex}{33}
\affiliation{\ROMAII}
\newcommand*{\MSU}{Skobeltsyn Institute of Nuclear Physics, Lomonosov Moscow State University, 119234 Moscow, Russia}
\newcommand*{\MSUindex}{34}
\affiliation{\MSU}
\newcommand*{\SCAROLINA}{University of South Carolina, Columbia, South Carolina 29208, USA}
\newcommand*{\SCAROLINAindex}{35}
\affiliation{\SCAROLINA}
\newcommand*{\TEMPLE}{Temple University,  Philadelphia, PA 19122, USA }
\newcommand*{\TEMPLEindex}{36}
\affiliation{\TEMPLE}
\newcommand*{\JLAB}{Thomas Jefferson National Accelerator Facility, Newport News, Virginia 23606, USA}
\newcommand*{\JLABindex}{37}
\affiliation{\JLAB}
\newcommand*{\UTFSM}{Universidad T\'{e}cnica Federico Santa Mar\'{i}a, Casilla 110-V Valpara\'{i}so, Chile}
\newcommand*{\UTFSMindex}{38}
\affiliation{\UTFSM}
\newcommand*{\EDINBURGH}{Edinburgh University, Edinburgh EH9 3JZ, United Kingdom}
\newcommand*{\EDINBURGHindex}{39}
\affiliation{\EDINBURGH}
\newcommand*{\GLASGOW}{University of Glasgow, Glasgow G12 8QQ, United Kingdom}
\newcommand*{\GLASGOWindex}{40}
\affiliation{\GLASGOW}
\newcommand*{\TUFTS}{Tufts University, Medford, MA 02155, USA}
\newcommand*{\TUFTSIndex}{41}
\affiliation{\TUFTS}
\newcommand*{\VT}{ Virginia Polytechnic Institute and State University, Virginia   24061-0435, USA}
\newcommand*{\VTindex}{42}
\affiliation{\VT}
\newcommand*{\VIRGINIA}{University of Virginia, Charlottesville, Virginia 22901, USA}
\newcommand*{\VIRGINIAindex}{43}
\affiliation{\VIRGINIA}
\newcommand*{\WM}{College of William and Mary, Williamsburg, Virginia 23187-8795, USA}
\newcommand*{\WMindex}{44}
\affiliation{\WM}
\newcommand*{\YEREVAN}{Yerevan Physics Institute, 375036 Yerevan, Armenia}
\newcommand*{\YEREVANindex}{45}
\affiliation{\YEREVAN}
\newcommand*{\NOWUK}{University of Kentucky, Lexington, Kentucky 40506, USA}
\newcommand*{\NOWISU}{Idaho State University, Pocatello, Idaho 83209, USA}
\newcommand*{\NOWJLAB}{Thomas Jefferson National Accelerator Facility, Newport News, Virginia 23606}
\newcommand*{\NOWINFNGE}{INFN, Sezione di Genova, 16146 Genova, Italy}

\author {M.C.~Kunkel}
\affiliation{\ODU}
\affiliation{\IKP}
\author {M.J.~Amaryan}
\thanks{Corresponding author; mamaryan@odu.edu}
\affiliation{\ODU}
\author {I.I.~Strakovsky}
\affiliation{\GWUI}
\author {J.~Ritman}
\affiliation{\BOCHUM}
\affiliation{\IKP}
\author{G.R.~Goldstein}
\affiliation{\TUFTS}
\author {K.P. ~Adhikari} 
\affiliation{\MISS}
\author {S~ Adhikari} 
\affiliation{\FIU}
\author{H.~Avakian}
\affiliation{\JLAB}
\author {J.~Ball} 
\affiliation{\SACLAY}
\author {I.~Balossino} 
\affiliation{\INFNFE}
% ...

\collaboration{The CLAS Collaboration}
%
\vskip 1.50in
\vskip 2in
\begin{abstract}
\centerline{\Large Abstract}
\vskip 0.2in
Exclusive photoproduction cross sections have been measured for the process $\gamma p \rightarrow p\pi^0(e^+e^-(\gamma))$ with the Dalitz decay final state  using tagged photon energies in the range of $E_{\gamma} = 1.275-5.425$~GeV. The complete angular distribution of the final state $\pi^0$, for the entire photon energy range up to large values of $t$ and $u$, has been measured for the first time.
The data obtained show that the cross section $d\sigma/dt$, at mid to large angles, decreases with energy as $s^{-6.89\pm 0.26} $. This is in agreement with the perturbative QCD quark counting rule prediction of $s^{-7} $. Paradoxically, the size of angular distribution of measured cross sections is greatly underestimated by the QCD based Generalized Parton Distribution  mechanism at highest available invariant energy $s=11$~GeV$^2$. At the same time, the Regge exchange based models for $\pi^0$ photoproduction are more consistent with experimental data.
\end{abstract}


\maketitle

In general, there are  properties of $\pi^0$ that make this particle very special for our understanding of Quantum Chromodynamics (QCD). To name a few: it is the lightest element of all visible hadronic matter in the Universe; according to its $q \bar q$  content the $\pi^0$ has a mass much less than one would expect from a constituent quark mass, $m \approx$~350~MeV and it has an extremely short life time, $\tau \approx 10^{-16}$~s. Its main decay mode,   $\pi^0\rightarrow \gamma \gamma$, with a branching ratio~$\approx 99\%$, played a crucial role in confirming the number of colors in QCD and in establishing the chiral anomaly in gauge theories. With all this being said,  the structure and properties of $\pi^0$ are not completely understood. 

One of the cleanest ways to obtain additional experimental information about the $\pi^0$ is high energy photoproduction on a proton, as the incoming electromagnetic wave is structureless, contrary to any hadronic probe. Even after decades of experimental efforts, precise data of the elementary reaction $\gamma  p \rightarrow p\pi^0 $, above the resonance region and at large values of all Mandelstam variables $s$, $t$, and $u$, are lacking.

At the interface between the crowded low energy resonance 
production regime and the smooth higher energy, small angle behavior, 
traditionally described by Regge poles~\cite{Ader:1967tqj}, lies a 
region in which hadronic duality interpolates the different excitation function
behavior. Exclusive $\pi$ photoproduction and $\pi$ nucleon elastic 
scattering show this duality in a semi-local sense through Finite Energy 
Sum Rules (FESR)~\cite{Armenian:1974xd}. The connection to QCD is more 
tenuous for on-shell photoproduction of pions at small scattering angles, 
but the quark content can become manifest through large fixed angle 
dimensional counting rules~\cite{Brodsky:1973kr}, \cite{Matveev:1972gb}, as well as being evident 
in semi-inclusive or exclusive electroproduction of pions, described 
through Transverse Momentum Distributions (TMDs)  and Generalized Parton 
Distributions (GPDs).


The Regge pole description of photoproduction amplitudes 
has a long and varied history. For $\pi^0$ and $\eta$ photoproduction, 
all applications rely on a set of known meson Regge poles. There are 
two allowed $t$-channel J$^{PC}$ quantum numbers, the odd-signature (odd spin) 
1$^{--}$ ~($\rho^0$, $\omega$) and the 1$^{+-}$~($b^0_1$, $h_1$) Reggeons. Regge cut amplitudes are 
incorporated into some models and are interpreted as rescattering of 
on-shell meson-nucleon amplitudes.  The phases between the different 
poles and cuts can be critical in determining the polarizations and the 
constructive or destructive interferences that can appear. Four distinct Regge models 
are considered here.

An early model developed by Goldstein and 
Owens~\cite{Goldstein:1973xn} has the exchange of leading Regge 
trajectories with appropriate $t$-channel quantum numbers along with 
Regge cuts generated via final state rescattering through Pomeron 
exchange. The Regge couplings to the nucleon were fixed by reference 
to electromagnetic form factors, SU(3)$_{flavor}$, and low energy 
nucleon-nucleon meson exchange potentials. At the time, the range of 
applicability was taken to be above the resonance region and $\mid 
t \mid \le 1.2 \, {\rm GeV}^2$, where $t$ is the squared 
four-momentum transfer. Here we will let the $t$ 
range extend to large values of $t$ in order to see the predicted cross 
section dips from the zeroes in the Regge residues. Because even signature partners ($A_2$, $f_2$) of the odd spin poles ($\rho$, $\omega$) lie on the same trajectories, the Regge residues are required to have zeroes to cancel the even (wrong) signature poles in the physical region - these extra zeroes are called nonsense wrong signature zeroes (NWSZ)~\cite{CHIU1971477}. While the dip near 
$t\approx -0.5$ GeV$^2$ is present in the $\pi^0$ cross section data, it is absent in the 
beam asymmetry, $\Sigma$, measurement for $\pi^{0}$ and $\eta$ 
photoproduction~\cite{AlGhoul:2017nbp}. This is not explained by the 
standard form of the NWSZ Regge residues. 
  
Quite recently, Mathieu {\it et al.}~\cite{Mathieu:2015eia} from the Joint Physics Analysis Center (JPAC)
(see also~\cite{Kashevarov:2017vyl}), used the same set of Regge poles, 
but a simplified form of only $\omega$ -Pomeron cuts. They show that 
daughter trajectories are not significant as an alternative to the 
Regge cuts. However, to reproduce the lack of $t\approx -0.5$ GeV$^2$ 
dip in $\eta$ photoproduction, they remove the standard wrong signature 
zero, i.e., the NWSZ.  Donnachie and Kalashnikova~\cite{Donnachie:2015jaa} 
have included $t$-channel $\rho^0$, $\omega$, and $b^0_1$ exchange, but not 
the $h_1$ Reggeon, all with different parameterizations from 
Ref.~\cite{Goldstein:1973xn}. They include $\omega, \rho \otimes {\rm 
Pomeron}$ cuts, as well as $\omega, \rho \otimes {\rm f}_2$ lower lying 
cuts, which help to fill in the wrong signature zeroes of the $\omega, 
\rho$ Regge pole residues. The model of Laget and collaborators~\cite{Laget:2005be} included 
$u$-channel baryon exchange, which dominate at backward angles, along with 
	elastic and inelastic unitarity cuts 
	to fill the intermediate 
	$t$ range. 
	With these ingredients, the model is expected to describe the full angular range ($\theta_{\pi}= 0 \to 180^{\circ}$), where $\theta_{\pi}$ is the pion polar production and in the c.m.~frame,
	while the other models are good for more limited ranges of $t$~\cite{Goldstein:1973xn,Mathieu:2015eia, Donnachie:2015jaa}. 
Here, we examine how Regge phenomenology works for the energy range of 2.8~GeV $< $ E$_\gamma$ $<$ 5.5~GeV.

In addition to Regge pole models, the introduction of the Handbag mechanism, developed by 
Kroll \textit{et al.}~\cite{Huang:2000kd}, has provided complementary 
possibilities for the interpretation of hard exclusive reactions. In 
this approach, the reaction is factorized into two parts, one quark 
from the incoming and one from the outgoing nucleon participate in the 
hard sub-process, which is calculable using Perturbative Quantum ChromoDynamics (pQCD). The soft part 
consists of all the other partons that are spectators and can be 
described in terms of GPDs~\cite{Ji:1996nm}.
The Handbag model applicability requires a hard scale, which, for meson 
photoproduction, is only provided by large transverse momentum, which 
corresponds to large angle production, roughly for 
$-0.6~\leq\cos \theta_{\pi}~\leq 0.6$.  Here, we examined how the Handbag 
model may extend to the $\gamma p\rightarrow p\pi^0$ case proposed in~\cite{Huang:2000kd}. The distribution amplitude for the 
quark+antiquark to $\pi^0$ is fixed by other phenomenology and 
leads to the strong suppression of the production cross section.


\begin{figure*}[htb]
	\centerline{
		\includegraphics[height=0.35\textwidth,width=0.5\textwidth]{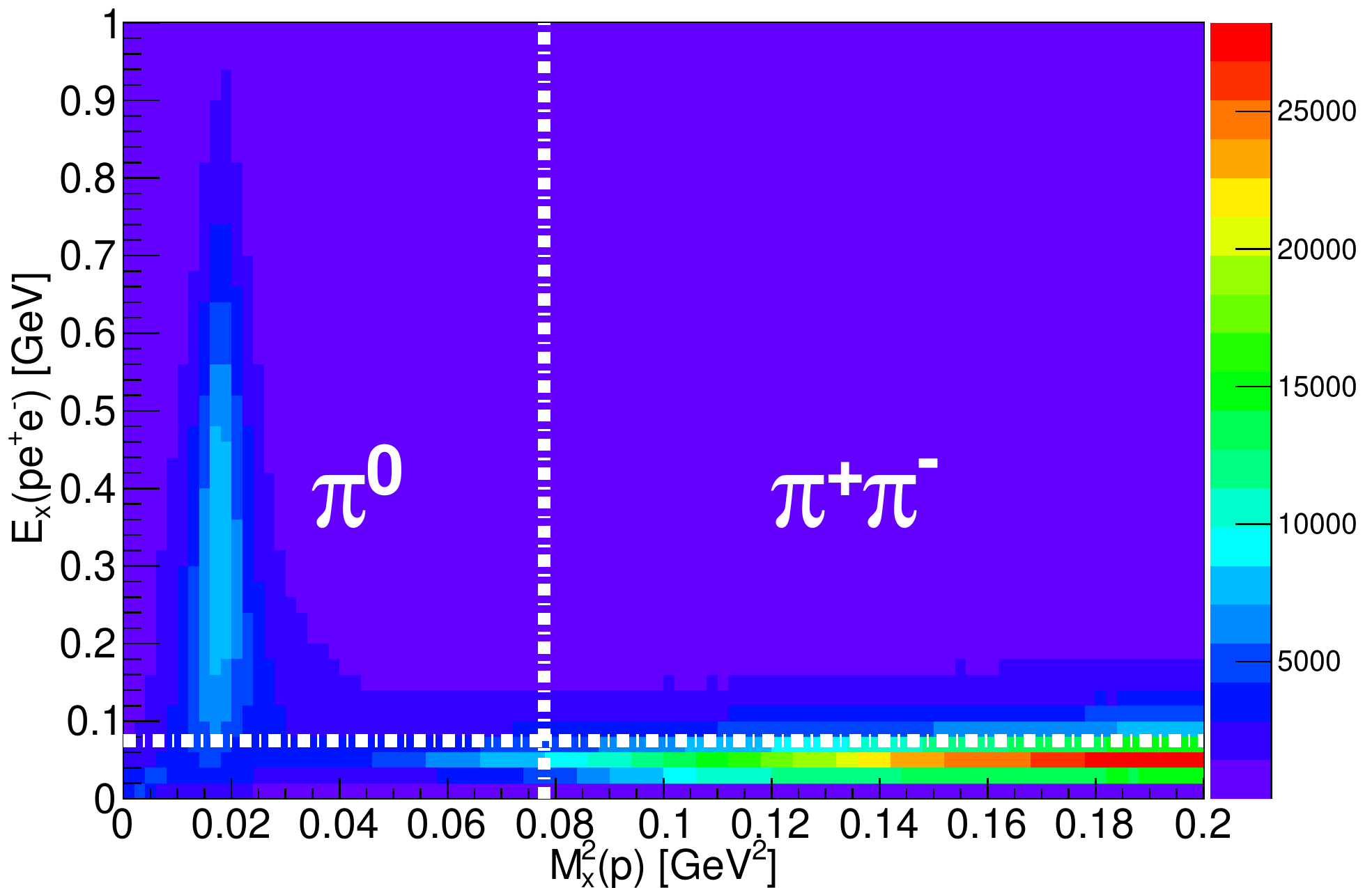}\hfill
		\includegraphics[height=0.35\textwidth,width=0.5\textwidth]{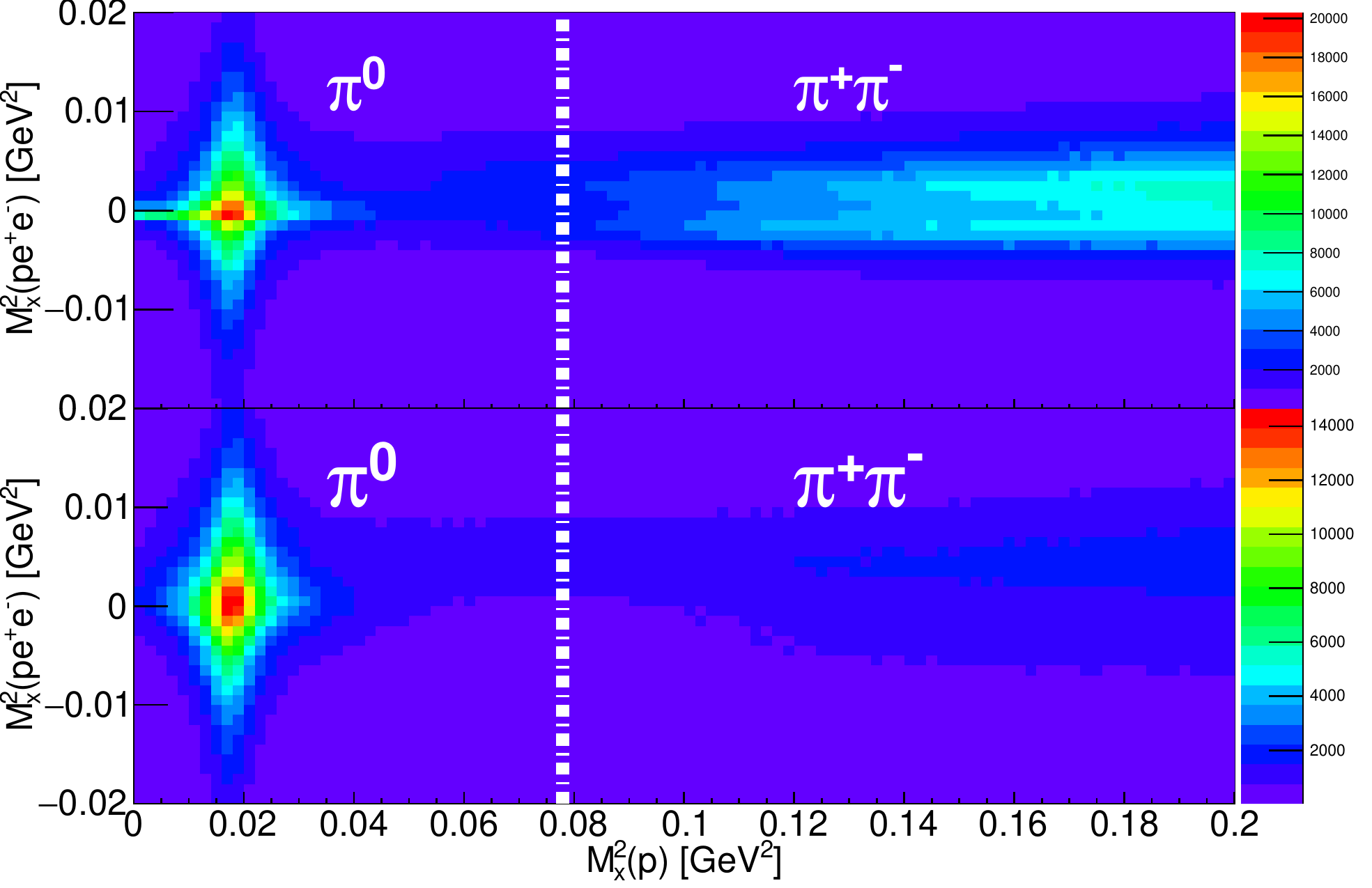}}
	
	\caption{(Color online)(left panel)Missing energy $\mathrm{E_X(pe^+e^-)}$ of all detected particles vs missing mass squared of the proton $  \mathrm{M_x^2(p)}$. (Right panel) Missing mass squared of all detected particles $\mathrm{M_x^2(pe^+e^-)}$ vs 
		missing mass squared of the proton $\mathrm{M_x^2(p)}$; before applying the
		the cut on missing energy, $\mathrm{E_X(pe^+e^-)} $,  (right-top panel), and 
		after applying the cut $\mathrm{E_X(pe^+e^-)} >$ 75~MeV  (right-bottom panel).
		The horizontal white dashed-dotted line depicted on the left
		panel illustrates the 75~MeV threshold used in this analysis.
		The vertical white dashed-dotted line depicts the kinematic
		threshold for $\pi^+\pi^-$ production.}
	\label{fig:sys}
\end{figure*}
\begin{figure}[!htb]
	\centerline{
		\includegraphics[height=0.35\textwidth,width=0.5\textwidth]{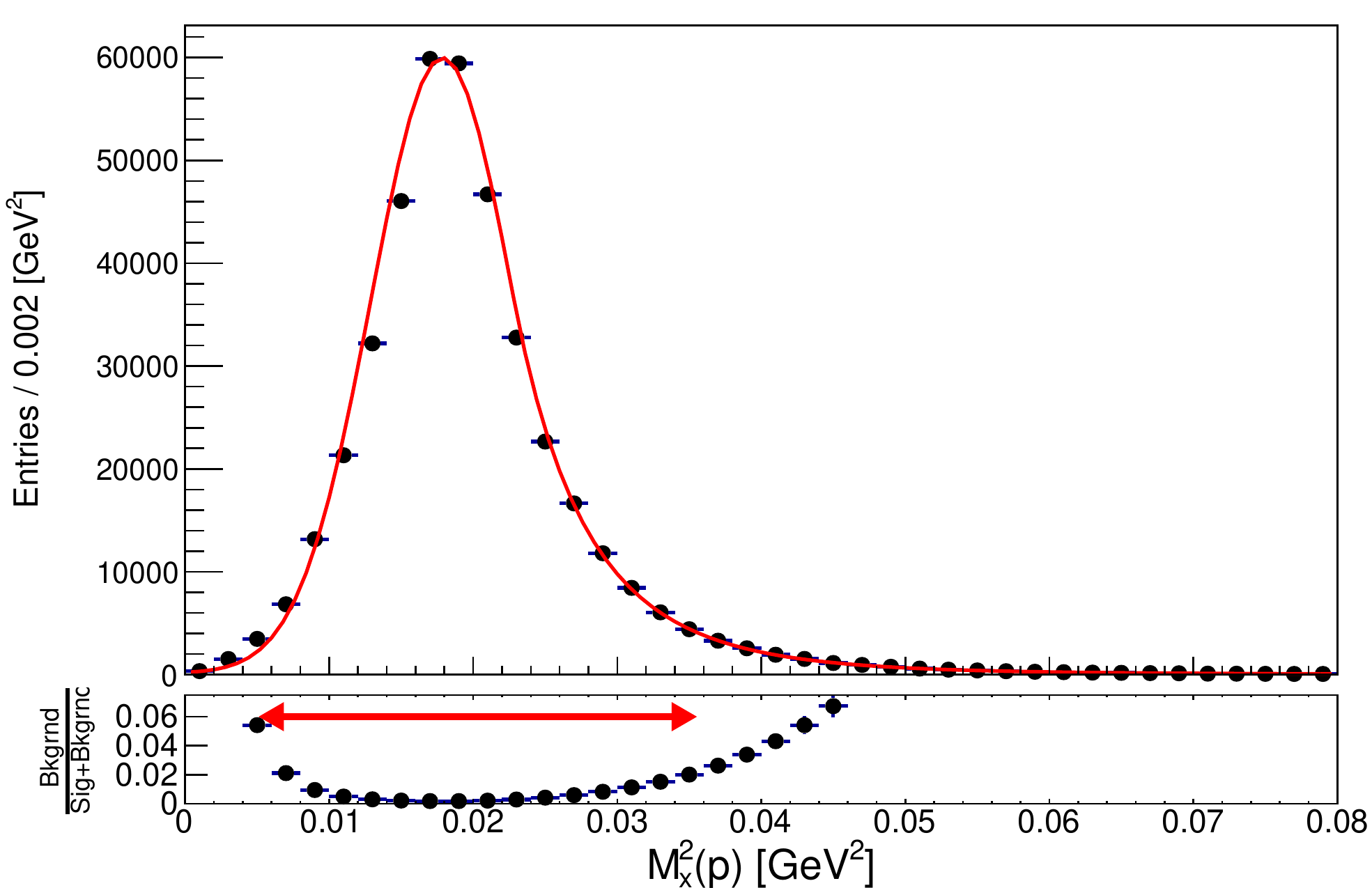}}
	
	\caption {(Color online) (top-panel) Peak of $\pi^0$ in the  proton missing mass squared for events 
		with $pe^+e^-(\gamma)$ in the final state.
		The red-solid line depicts the fit function (signal+background).
		(bottom-panel) Relative contributions of $\frac{\mathrm{Background}}
		{\mathrm{Signal + Background}}$. The red arrow indicates the cut
		placed on the $M_x^2(p)$ distribution to select \pizT events.}
	\label{fig:pi0_peak}
\end{figure}
Binary reactions in QCD with large momentum transfer 
occur via gluon and quark exchanges between the colliding particles. The 
constituent counting rules ~\cite{Brodsky:1973kr} \cite{Matveev:1972gb}
provide a simple recipe to predict the energy dependence of the 
differential cross sections of two-body reactions at large angles 
when the ratio $t/s$ is finite and is kept constant.  The lightest meson 
photoproduction was examined in terms of these counting 
rules~\cite{Anderson:1976ph,Jenkins:1995bk,Zhu:2002su,Chen:2009sda,
Kong:2015yzn}. As was first observed at SLAC by 
Anderson \textit{et al.}~\cite{Anderson:1976ph}, the reaction $\gamma p\rightarrow n \pi^+$ 
shows agreement with constituent counting rules that predict the 
cross section should vary as $s^{-7}$. The 
agreement extends down to $s$ = 6~GeV$^2$ where baryon resonances are 
still playing a role.  Here, we examined how applicable the counting rule is 
for $\gamma p\rightarrow  p\pi^0$ up to $s$ = 11~GeV$^2$. 

An earlier, untagged bremsstrahlung, measurements of $\gamma p\rightarrow p\pi^0$, for $2~\leq E_{\gamma} \leq 
18$~GeV (1964 -- 1979) provided 451 data points for differential cross section $d\sigma/dt$~\cite{brem}, have very large systematic 
uncertainties and do not have sufficient accuracy to perform 
comprehensive phenomenological analyses.  A previous CLAS measurement of $\gamma p\rightarrow p\pi^0$, for $2.0~\leq E_{\gamma} \leq 2.9$~GeV, has an overall systematic uncertainty of 5\% but only provided 164 data points for differential cross section $d\sigma/dt$~\cite{Dugger:2007bt}.

The results described here are the first to allow a detailed analysis, bridging the nucleon resonance and high energy regions over a wide angular range, of exclusive pion photoproduction. By significantly extending the database they facilitate
the examination
%
		of the resonance, ``Regge", and wide angle QCD regimes of phenomenology. The 
		broad range of c.m. energy, $\sqrt{s}$, is particularly helpful in 
		sorting out the phenomenology associated with both Regge and QCD-based 
		models of the nucleon~\cite{Kroll:2017hym}.

In this work, we provide a large set of differential 
cross section values from $E_{\gamma} = 1.25-5.55$~GeV in laboratory photon 
energy, corresponding to a range of c.m. energies, $W$ = 1.81 -- 
3.33~GeV.  We have compared the Regge pole, the Handbag, and the 
constituent counting rule phenomenology with the new CLAS experimental 
information on $d\sigma/dt$ for the $\gamma p\rightarrow  p\pi^0$ 
reaction above the ``resonance" regime. As will be seen, this data 
set quadruples the world  database for $\pi^0$ photoproduction above $E_{\gamma} =$ 2~GeV and 
constrains the high energy phenomenology well with a previous CLAS 
measurement~\cite{Dugger:2007bt}.


The experiment was performed during March-June, 2008
with the CLAS detector at Jefferson Laboratory~\cite{meck} using a energy-tagged photon beam produced by 
bremsstrahlung from a 5.72~GeV electron beam provided by the CEBAF 
accelerator, which impinged upon a liquid hydrogen target,
and was designated with the name $g12$. 
The experimental details are given in Ref.~\cite{g12}. The reaction 
of interest is the photoproduction of neutral pions on a hydrogen 
target $\gamma p\rightarrow p\pi^0$, 
where the neutral pions decay into an $e^+e^-\gamma$ final state either due to external conversion, $\pi^0 \rightarrow\gamma\gamma 
\rightarrow e^+e^-\gamma$ or via Dalitz decay $\pi^0
\rightarrow\gamma^\ast\gamma\rightarrow e^+e^-\gamma$. Running the 
experiment at high beam current was possible due to the final state 
containing three charged tracks, $p$, $e^+$, $e^-$, as opposed to single 
prong charged track detection which impose limitations due to trigger 
and data acquisition restrictions.


Particle identification for the experiment was based on $\beta$ vs. momentum$\times$charge. 
Lepton identification was based on a kinematic constraint to the $\pi^0$ mass. 
Once the data was skimmed for $p$, $\pi^+$, and $\pi^-$ tracks, 
all particles that were $\pi^+$, $\pi^-$ were tentatively assigned 
to be electrons or positrons based on their charge (for details, 
see Ref.~\cite{Kunkel,KunkelPhD}). After particle selection, standard $g12$ 
calibration, fiducial cuts and timing cuts were applied 
in the analysis~\cite{g12}.

\begin{figure*}[htb!]
	\centerline{
		\includegraphics[height=0.65\textwidth, angle=90]{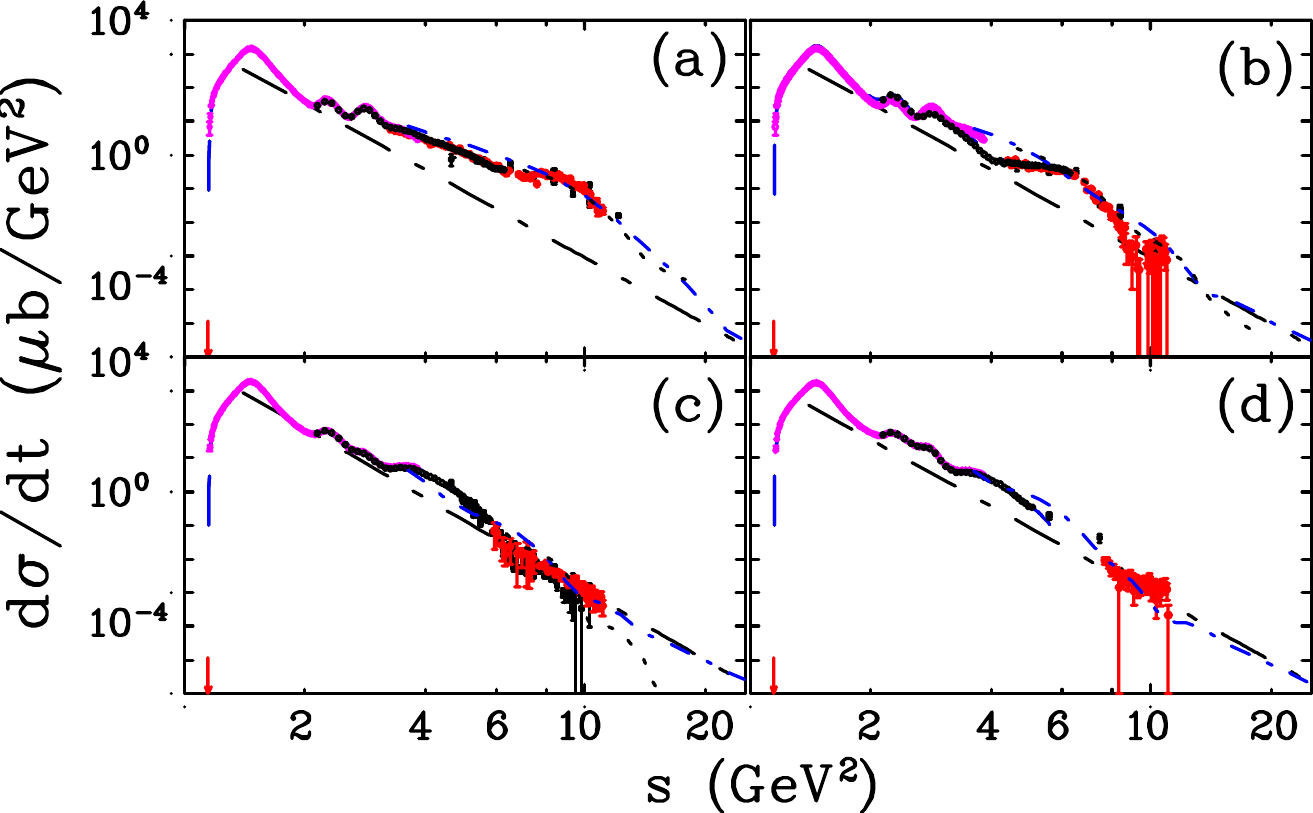}}
	\caption {(Color online) Differential cross section $d \sigma/dt$ of the reaction
		$\gamma p\rightarrow\  p\pi^0 $ at polar angles 
		of (a) 50$^\circ$, (b) 70$^\circ$, 
		(c) 90$^\circ$, and (d) 
		110$^\circ$ in the c.m. frame as a function of c.m. energy 
		squared, $s$. The red filled circles are the current $g12$ 
		CLAS data. The recent tagged photon data are from previous 
		CLAS Collaboration 
		measurements~\protect\cite{Dugger:2007bt} (black open 
		circles) and the A2 Collaboration at 
		MAMI~\protect\cite{Adlarson:2015byy} 
		(magenta open diamonds with crosses), while the black filled 
		squares are data from old bremsstrahlung measurements above 
		$E_{\gamma}$ = 2~GeV~\protect\cite{brem}. The plotted uncertainties are 
		statistical.  
		The blue dashed line corresponds to the SAID PWA 
		PR15 solution (no new CLAS $g12$ data are used 
		for the fit)~\protect\cite{Adlarson:2015byy}.  The black dot-dashed 
		lines are plotted as the best fit result of the power function $s^{-n}$, with $n = 6.89\pm$0.26, for the spectrum at 
		90$^\circ$. The pion production threshold is shown as a vertical 
		red arrow. The Regge results~\protect\cite{Goldstein:1973xn,
			Laget:2005be} are given by the black dotted line and the blue short dash-dotted line, 
		respectively.} \label{fig:scaling}
\end{figure*}
Different kinematic fits were employed to cleanly identify the 
$\gamma p\rightarrow pe^+e^-(\gamma)$ reaction. They were applied 
to filter background from misidentified double pion production to 
the single $\pi^0$ production, to constrain the missing mass of entire 
final state to a missing photon and to ensure that the fit to the missing 
photon constrained the squared invariant 
mass of $e^+e^-(\gamma)$=m$^2_{\pi^0}$. The values of the confidence levels cuts employed was 
determined using the statistical significance to get the best signal/background ratio.
The confidence levels for each constraint were consistent 
between the $g12$ data and Monte-Carlo simulations. 
Monte-Carlo generation was performed using the PLUTO++ package 
developed for the HADES Collaboration~\cite{PLUTO}.

The remainder of the background was attributed to $\pi^+\pi^-$
events. To reduce the background further, a comparison of the 
missing mass squared off the proton, $\mathrm{M_x^2(p) =(P_\gamma + P_p -P^{'}_p)^2}$, in terms of the four-momenta of the incoming photon, target proton, and final state proton, respectively, and the missing energy of detected system, $\mathrm{E_X(pe^+e^-)}= E_\gamma + E_p  -  E^{'}_p - E_{e^{+}} - E_{e^{-}} $, was performed, see Fig.~\ref{fig:sys}. This 
comparison revealed that the majority of the $\pi^+\pi^-$ background 
has missing energy less than 75~MeV. To eliminate this background 
all events with a missing energy less than 75~MeV were removed.

The distribution of the proton missing mass squared for events with 
$pe^+e^-(\gamma)$ in the final state is shown in Fig.~\ref{fig:pi0_peak}. 
A fit was performed with the Crystal Ball function~\cite{Oreglia:1980cs,
Skwarnicki:1986xj} for the signal, plus a 3rd order polynomial function 
for the background. The total signal+background fit is shown by the red solid 
line. The fit resulted in $M_{\pi^0}^2$ = 0.0179~GeV$^2$ with a Gaussian width
$\sigma$=0.0049~GeV$^2$. To select \pizT events, an asymmetric cut about the measured value was placed 
in the range $0.0056 $~GeV$^2 \le  M_x^2(p) 
\le 0.035$~GeV$^2$. This cut range can be seen as the arrow in the bottom 
panel of Fig.~\ref{fig:pi0_peak} along with the ratio of background 
events to the total number of events. As shown in Fig.~\ref{fig:pi0_peak}, 
the event selection strategy for this analysis led to a 
negligible integrated background estimated to be no more than $1.05\%$.

The total systematic uncertainty varied between 9\% and 12\% as a function of energy. The individual contributions came from particle efficiency, sector-to-sector efficiency, 
flux determination, missing energy cut, the kinematic fitting probabilities, 
target length, branching ratio, fiducial cut, and the $z$-vertex cut.
The largest contributions to the systematic uncertainties 
were the sector-to-sector (4.4 -- 7.1\%), flux determination (5.7\%),
and the cut on the 1-C pull probability (1.6 -- 6.1\%). All systematic 
uncertainties and their determinations are described in Ref.~\cite{Kunkel}.



As it was mentioned above there are two subprocesses that may led to the same final state $\pi^0 \rightarrow e^+e^-\gamma$. Both subprocesses
were simulated in the Monte Carlo with their corresponding branching ratios and used to obtain cross sections from experimentally observed yield of neutral pions.

\begin{figure*}[htb!]
\centerline{
        \includegraphics[width=2.2in, angle=90]{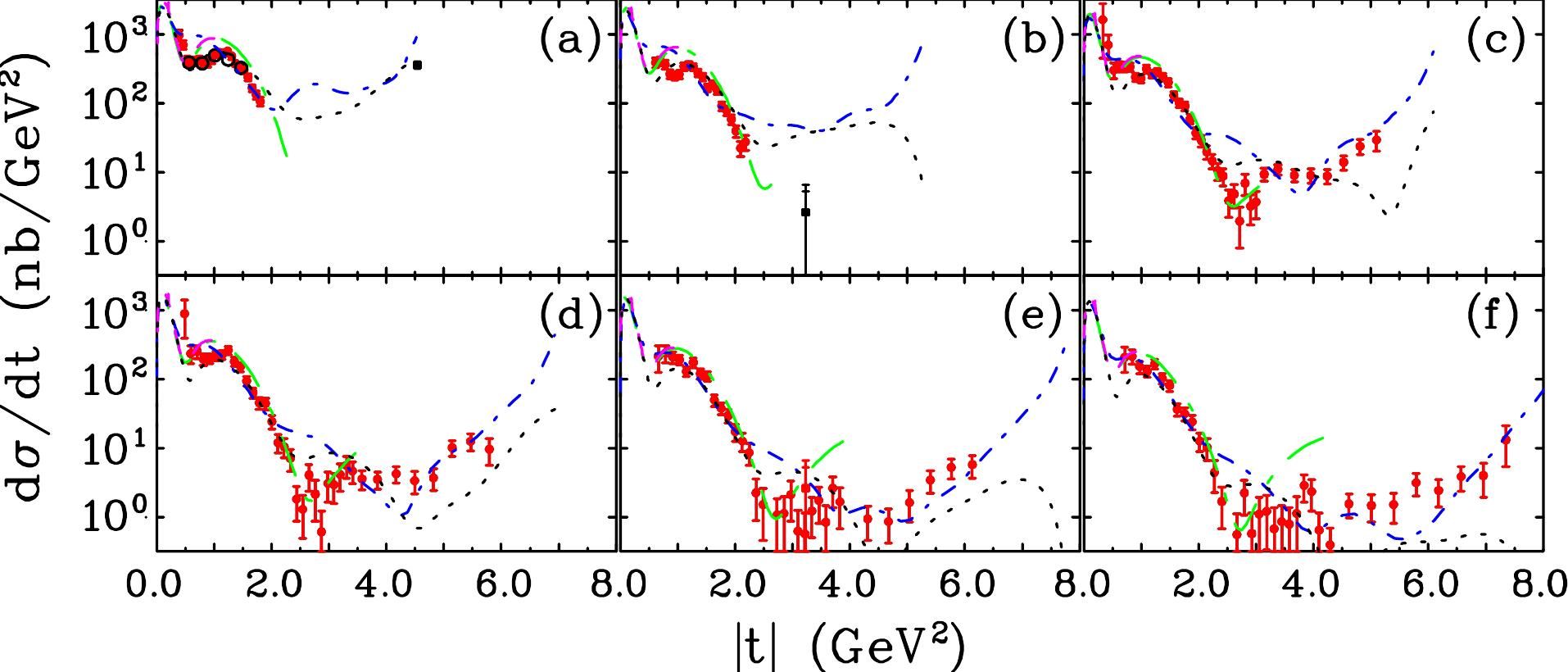}}

        \caption {(Color online) Samples of the $\pi^0$ photoproduction 
	cross section, $d\sigma/dt$, off the proton versus $|t|$ 
	above ``resonance" regime.  
	(a) $E_{\gamma}$ = 2825~MeV and $W$ = 2490~MeV, 
	(b) $E_{\gamma} $ = 3225~MeV and $W$ = 2635~MeV,
	(c) $E_{\gamma}$ = 3675~MeV and $W$ = 2790~MeV, 
	(d) $E_{\gamma}$ = 4125~MeV and $W$ = 2940~MeV,
	(e) $E_{\gamma}$ = 4575~MeV and $W$ = 3080~MeV, and
	(f) $E_{\gamma}$ = 4875~MeV and $W$ = 3170~MeV.
	Tagged experimental data are from the current CLAS $g12$ measurements (red 
	filled circles) and a previous CLAS 
	measurement~\protect\cite{Dugger:2007bt} (black open circles). 
	The plotted points from previously published bremsstrahlung 
	experimental data above $E_{\gamma}$ = 2~GeV~\protect\cite{brem} (black 
	filled squares) are those data points within $\Delta E_{\gamma} = 
	\pm$3~MeV of the photon energy in the laboratory system 
	indicated on each panel. The uncertainties plotted are only 
	statistical. 
	Regge results~\protect\cite{Goldstein:1973xn,
	Mathieu:2015eia,Donnachie:2015jaa,Laget:2005be} are given by black dotted line, 
	green dot-dashed line, magenta long 
	dashed line, and blue short dash-dotted, respectively.} 
	\label{fig:t_data}
\end{figure*}

The new CLAS high statistics $\gamma p \to \pi^0 p$ cross sections from
this analysis are compared in Figs.~\ref{fig:scaling}
and \ref{fig:t_data} with data from previous CLAS~\cite{Dugger:2007bt},  untagged bremsstrahlung data of DESY, Cambridge
Electron Accelerator (CEA), and SLAC, and Electron Synchrotron at
Cornell Univ. measurements~\cite{brem}, as
well as lower c.m. energy measurements by A2 Collaboration at MAMI~\protect\cite{Adlarson:2015byy} with tagged photon beam.
%
The overall agreement is good,
particularly with the previous CLAS data.

\begin{figure}[htb!]
	\centerline{
		\includegraphics[width=2.0in, angle=90]{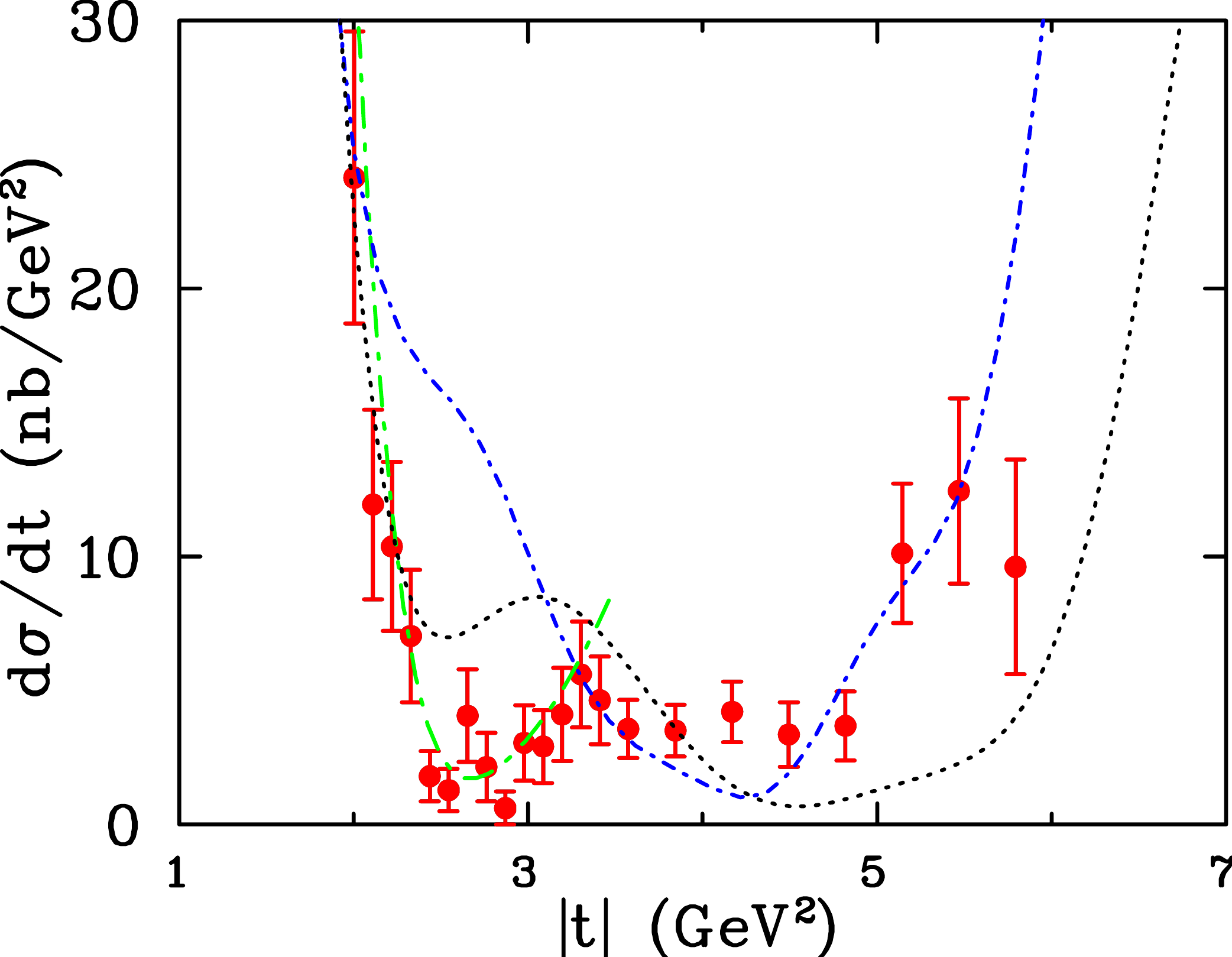}}
	\caption {(Color online) Sample of the $\pi^0$ photoproduction 
		cross section, $d\sigma/dt$, off the proton versus $|t|$ 
		above ``resonance" regime at $E_{\gamma}$ = 4125~MeV and $W$ = 2940~MeV.
		The theoretical curves for the Regge 
		fits are the same as in Fig.~\protect\ref{fig:t_data}. }
	\label{fig:t_dataB}
\end{figure}
\begin{figure}[htb!]
	\centerline{
		\includegraphics[width=2.0in, angle=90]{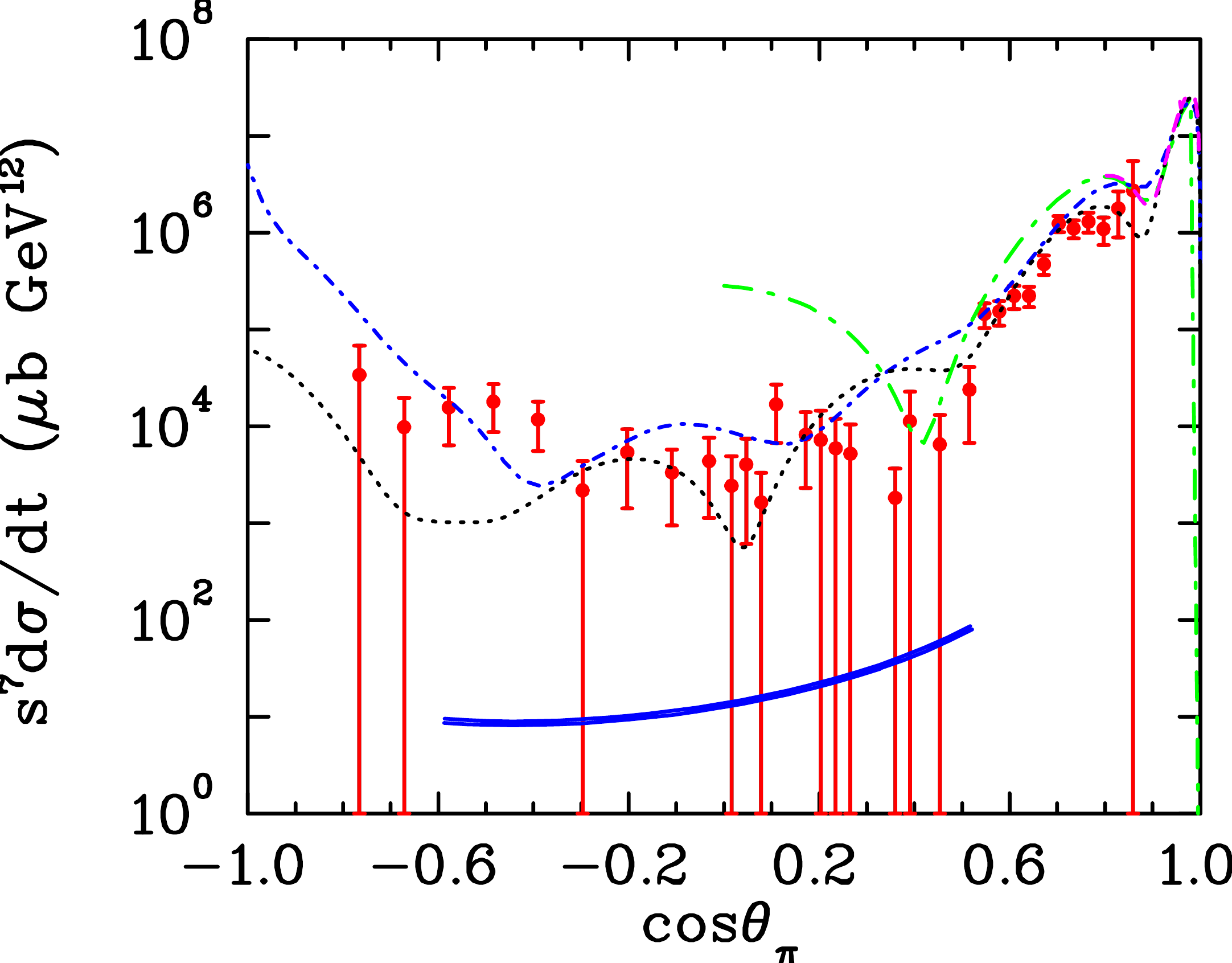}}
	
	\caption {(Color online) Differential cross section 
		of $\pi^0$ photoproduction. The CLAS experimental data 
		at $s $~= 11GeV$^2$ are from the current experiment (red 
		filled circles). The theoretical curves for the Regge 
		fits are the same as in Fig.~\protect\ref{fig:t_data} 
		and the Handbag model by Kroll~\textit{et 
			al.}~\protect\cite{Huang:2000kd} (blue double solid 
		line).} \label{fig:kroll}
\end{figure}

At higher energies (above $s\sim$ 6~GeV$^2$) and large c.m. angles 
($\theta_{\pi}\geq$ 90$^\circ$), the results are consistent with 
the $s^{-7}$ scaling, at fixed $t/s$, as expected from the 
constituent counting rule~\cite{Brodsky:1973kr}. 
The black dash-dotted line at 90$^\circ$ (Fig.~\ref{fig:scaling}) 
is a result of the fit of new CLAS $g12$ data only, performed with a 
power function $\sim s^{-n}$, leading to $n = 6.89\pm$0.26.  
Structures observed at 50$^\circ$ and 70$^\circ$ up to 
$s\sim$11~GeV$^2$ indicate that the constituent 
counting rule requires higher energies and higher $|t|$ before it 
can provide a complete description.
In Figs.~\ref{fig:t_data},~\ref{fig:t_dataB} and \ref{fig:kroll}, the 
$d\sigma/dt$ results are shown along with predictions from 
Regge pole and cut~\cite{Goldstein:1973xn,Laget:2005be,
Mathieu:2015eia,Donnachie:2015jaa} models and the 
Handbag~\cite{Huang:2000kd} model. 

Overall, the Regge approximation 
becomes less applicable below $E_{\gamma}$ = 3~GeV (Fig.~\ref{fig:t_data}). 
Below $|t|\sim$1.0~GeV$^2$ there is a small difference between 
different Regge approaches.    
Note that some small dips start to appear around $|t| \sim 0.5$~GeV$^2$
~($\cos \theta_{\pi} = 0.6-0.8$) where the Regge models predict a dip.  
Prior to this measurement there was no indication of these dips.
Note that the Regge amplitudes impose non-negligible constraints when continued down to the 
``resonance" region.
Our data show another visible dip above $E_{\gamma}$ = 3.6~GeV at around $|t|\sim$2.6~GeV$^2$ and possible manifestation of another``possible new structure'' around $|t|\sim$5~GeV$^2$ for $E_{\gamma} >$ 4.1~GeV, where the Regge models~\cite{Goldstein:1973xn,
Laget:2005be,Donnachie:2015jaa} predict wrong signature zeroes, 
this is where the Regge trajectories cross negative even integers. 
For the dominant vector meson Regge poles, these dips should appear 
at approximately $-t=0.6, \, 3.0, \, 5.0$~GeV$^2$,  which agrees 
with the data. For a better visibility of these dips, as an example, a magnified version of Fig~\ref{fig:t_data}, for $E_\gamma$ = 4.125~GeV, is shown in Fig~\ref{fig:t_dataB}. 
The dip at about $|t|\sim$5.0~GeV$^2$ is best modeled by~\cite{Goldstein:1973xn}. 
The description of the $\pi^0$ photoproduction 
cross sections at largest $\vert t \vert$ requires improving the 
Regge model by including, for instance, additional exchange mechanisms.
\FloatBarrier

Fig.~\ref{fig:kroll} shows that the new CLAS data are orders of 
magnitude higher than the Handbag model prediction~\cite{Huang:2000kd} for $\pi^0$ 
photoproduction below $s$ = 11~GeV$^2$.


In this experiment a novel approach was employed based on the $\pi^{0}$ Dalitz decay 
mode. Although this decay mode has a branching fraction of only about 1\%, 
the enhanced event trigger selectivity enabled the figure of merit to be 
sufficiently high in order to extend the existing world measurements into 
an essentially unmeasured {\it terra incognita} domain.
Through the experiments described above, an extensive and precise 
data set (2030 data points) on the differential cross section for 
$\pi^0$ photoproduction from the proton has been obtained 
for the first time, except for a few points from previous measurements, 
over the range of $1.81~\leq W\leq 3.33$~GeV. 

Measurements were 
performed in the reaction $\gamma p\rightarrow pe^+e^-(\gamma)$ 
using a tagged photon beam spanning the energy interval covered 
by the ``resonance" and ``Regge" regimes.
The measurements obtained here have been compared to existing 
data. The overall agreement is good, while the data provided 
here quadrupled the world bremsstrahlung database above $E_{\gamma}$ = 
2~GeV and covered the previous reported energies with finer 
resolution. This new and greatly expanded set of data provides strong confirmation of the basic features of models based on Regge poles and cuts. There is enough precision to discriminate among the distinct components of those models. Guided by this data, extensions of models and improved parameterization is now possible. From another perspective, the wide angle data agree with the pQCD based constituent counting rules. Yet a significant paradox now appears: the wide angle data disagree - by orders of magnitude - with a handbag model that combines pQCD with the soft region represented by GPDs. This is an important result that needs to be better understood.

  

We thank Stanley Brodsky, 
Alexander Donnachie, Peter Kroll, Vincent Mathieu, 
and Anatoly Radyushkin for discussions of our measurements. We 
would like to acknowledge the outstanding efforts of the staff of 
the Accelerator and the Physics Divisions at Jefferson Lab that 
made the experiment possible.  This work was supported in part by 
the Italian Istituto Nazionale di Fisica Nucleare, the French 
Centre National de la Recherche Scientifique and Commissariat \`a 
l'Energie Atomique, the United Kingdom's Science and Technology 
Facilities Council (STFC), the U.~S. DOE and NSF, and the National Research Foundation of Korea. The Southeastern Universities 
Research Association (SURA) operates the Thomas Jefferson National 
Accelerator Facility for the US DOE under contract DEAC05--84ER40150.

\end{document}